\documentclass[preprint,showpacs,preprintnumbers,amsmath,amssymb] {revtex4}
\usepackage{graphicx}
\usepackage{dcolumn}
\usepackage{bm}
\newcommand{\bee}{\begin{eqnarray}}
\newcommand{\eend}{\end{eqnarray}}
\newcommand{\rmd}{{\rm d}}
\newcommand{\rme}{{\rm e}}

\newcommand{\bea}{\begin{eqnarray}}
\newcommand{\eea}{\end{eqnarray}}

\begin{document}

\title{Effective Lagrangian in nonlinear electrodynamics and its properties of causality and unitarity}

\author{Anatoly E. Shabad$^1$ and Vladimir V. Usov$^2$}
\affiliation{ $^1$ P.N. Lebedev Physics Institute, Moscow 117924,
Russia\\
$^2$Center for Astrophysics, Weizmann Institute, Rehovot 76100,
Israel}
\begin{abstract} In nonlinear electrodynamics, by implementing the causality
 principle as the requirement  that the group velocity of elementary excitations
 over a background field should not exceed the speed of light in the vacuum $c=1$,
 and the unitarity principle
 as the requirement that the residue of the propagator should be
 nonnegative, we  establish the positive convexity of the
 effective Lagrangian on the class of constant fields, also
 the positivity of all characteristic dielectric and magnetic permittivity
 constants that are derivatives of the effective Lagrangian with
 respect to the field invariants.
 Violation of the general principles by the one-loop approximation
 in QED at exponentially large magnetic
field is analyzed  resulting in complex energy ghosts that signal
the instability of the magnetized vacuum. Superluminal excitations
(tachyons) appear, too, but for the magnetic field exceeding its
instability threshold. Also other popular Lagrangians are tested to
establish that the ones leading to spontaneous vacuum magnetization
possess wrong convexity.
\end{abstract}
 \maketitle
\section{Introduction} The effective action that is
defined as the Legendre transform of the generating functional of
the Green functions \cite{weinberg}  and, in its turn, is itself a
generating functional of the (one-particle-irreducible) vertices
makes a basic quantity in  quantum field theory. This is a
c-numerical functional of fields and their derivatives, a knowledge
of which is meant to supply one with the final solution to the
theory. For this reason it seems important to see, how the most
fundamental principles manifest themselves as some general
properties of the effective action to be respected by model- or
approximation-dependent calculations, and whose violation might
signal important inconsistencies in the theory underlying these
calculations. Such inconsistencies may show themselves first of all
as ghosts and tachyons, that play an important role \cite{aref'eva}
in cosmological speculations about  forming the $\Lambda$-term and
dark energy using a scalar (Higgs) field yet to be discovered in the
coming experiments on the Large Hadronic Collider.

It is stated \cite{weinberg} basing on a formal continual integral
representation for the propagator that, when the effective action
$\Gamma(\phi)$ of a scalar field with mass $m$ is considered, its
second variational derivative
$\Sigma(x-y|\phi_0)=\left.\delta^2\Gamma/\delta\phi
(x)\delta\phi(y)\right|_{\phi=\phi_0}$ calculated at the constant
background value of this field, $\phi(x)=\phi_0,$ \textit{i.e.} the
mass operator against this background, is a nonpositive quantity,
$\Sigma\leq 0$. In other words, the effective Lagrangian is expected
-- to the extent that this formal property survives  perturbative or
other approximate calculations -- to be a concave = negatively
convex function (while the effective potential to be a (positively)
convex function) of a constant scalar field. On the other hand, the
same statement may be considered as the one directly prescribed by
the causality principle. Indeed, the spectral curve of small
excitations over the constant field background, $k_0=\sqrt{{\bf
k}^2+m^2-\Sigma (k)},$ where $k=(k_0,{\bf k})$ is the (4-momentum)
variable, Fourier-conjugate to the 4-coordinate difference $x-y$,
satisfies the causal propagation condition reading that its group
velocity should not exceed unity, the absolute speed limit for any
signal, $|\partial k_0/\partial {\bf k}|=|{\bf k}|/k_0 \leq 1$ for
any nonnegative  mass squared $m^2\geq 0$, provided, again, that
$\Sigma\leq 0$.

The case under our consideration here is much less trivial as we
deal not with a massive scalar, but with a massless vector gauge
field. The results apply, first of all, to nonlinear
electrodynamics, but also to (Abelian sector of) nonAbelian theory.
(Nonlinear electrodynamic models, the same as scalar ones, are also
considered for cosmological purposes \cite{novello-1} with the
advantage that instead of the scalar field, uncertain to be
physically identified, only well established electromagnetic field
is involved.)

We are going to demonstrate that the requirement of the causal
propagation of elementary excitations over the vacuum occupied by a
background field with a constant and homogeneous field strength,
supplemented by the requirements of translation-, Lorentz-, gauge-,
P- and C- invariances and unitarity has a direct impact on the
effective Lagrangian. For  the case -- which is general for
electromagnetic field, but special for a nonabelian field -- where
the Lagrangian depends on gauge-invariant combinations (field
strengthes) $F_{\alpha\beta}(z)=\partial_\alpha
A_\beta(z)-\partial_\beta A_\alpha(z)$ of the background field
potentials $A_\alpha(z)$, we make sure that the above requirements
are expressed as certain inequalities to be obeyed by the  first and
second derivatives of the effective Lagrangian with respect to the
two {\em field invariants} $\mathfrak{F}=\frac
1{4}F_{\rho\sigma}F_{\rho\sigma}=\frac 1{2}(B^2-E^2)$ and
$\mathfrak{G}=\frac
1{4}F_{\rho\sigma}\tilde{F}_{\rho\sigma}=(\bf{EB}),$ where $\bf E$
and $\bf B$ are background electric and magnetic fields,
respectively, and the dual field tensor is defined as
$\tilde{F}_{\rho\sigma}=\frac
1{2}\epsilon_{\rho\sigma\lambda\varkappa}F_{\lambda\varkappa}$,
where the completely antisymmetric unit tensor is defined in such a
way that $\epsilon_{1230}=1$. More specifically, we demonstrate that
it is a convex function with respect to the both variables
$\mathfrak{F,G}$ for any constant value of $\mathfrak{F}\gtrless 0$
and $\mathfrak{G}=0$. Note, the opposite sign of  convexity as
compared to the scalar field mentioned above.

In Section II  model- and approximation-independent study is
undertaken.

In Subsection A we remind the general diagonal representation of the
polarization operator and photon Green function in terms of its
eigenvectors and eigenvalues, obtained for arbitrary values of the
momentum $k$ and for nonzero constant field invariants
$\mathfrak{F,G}$ in \cite{batalin}, and refer to our previous work
\cite{newphysrev} where limitations on the location of dispersion
curves, imposed by demanding that the group velocity of the vacuum
excitations be less than/or equal to unity were established for the
general case of nonvanishing invariants $\mathfrak{F}$ and
$\mathfrak{G}$.

The unitarity requirement that the residue of the Green function in
the pole, corresponding to
 the mass shell of the elementary excitation, be nonnegative (completeness of
the set of states with nonnegative norm), is formulated.

In Subsection B we confine ourselves to the infrared asymptotic
behavior $k_\mu\rightarrow0$ of the polarization operator, in which
case its eigenvalues can be expressed in terms of first and second
derivatives of the effective Lagrangian with respect to the field
invariants $\mathfrak{F,G}$ when these are coordinate-independent.
Massless dispersion curves are explicitly found in terms of these
derivatives for the "magnetic-like" case $\mathfrak{F}>0$,
$\mathfrak{G}=0.$ The restrictions of Subsection A, now supplemented
with the unitarity requirement,   actualize as a number of
inequalities, to be satisfied by these derivatives. They mean, in
particular, that the effective Lagrangian is a (positively) convex
function of the field invariants in the point $\mathfrak{G}=0.$ We
reveal the physical sense of the quantities subject to these
inequalities as dielectric and magnetic permeabilities responsible
for polarizing small static charges and currents of special
configurations (There is no universal linear response function able
to cover every configuration, which is typical of an anisotropic
medium, to which class the magnetized vacuum belongs). In Subsection
C the inequalities of Subsection B are extended to include also the
"electriclike" background field $\mathfrak{F}<0$, $\mathfrak{G}=0,$
so in the end the whole axis of the variable $\mathfrak{F}$ is
included into result.

In Subsection D we write the (quadratic in the photon field )
contribution of the polarization operator into effective Lagrangian,
which is local in the infrared limit and presents the Lagrangian for
small, slow, long-wave perturbations of the background field
(infrared photons). This enables to define their energy-momentum
tensor via the Noether theorem. Once this is done, it becomes
possible to derive inequalities on the derivatives of the effective
Lagrangian basing on alternative pair of general requirements,
namely, the Weak Energy Condition  and Dominant Energy Condition of
Hawking and Ellis \cite {hawking} that are   positivity of the
energy density and non-spacelikeness of the energy-momentum flux
vector. We demonstrate that within our context the Dominant Energy
Condition is equivalent to restrictedness of the group velocity,
while the two alternative conditions together lead to a set of
inequalities, to which  the derivatives of the effective Lagrangian
are subjected, that do not contradict to the ones deduced in
Subsection B, but cannot be reduced to them. This implies that the
Weak Energy Condition is weaker than the positiveness of the residue
of the photon propagator exploited in Subsection B.

In Section III we test whether the properties resulting from the
general principles as derived in Section II are obeyed within
certain approximations and models. First we study the
Euler-Heisenberg one-loop effective Lagrangian of Quantum
Electrodynamics (Subsection A) and  the Lagrangian of Born and
Infeld (Subsection B) to establish that the latter perfectly
satisfies all of the above properties. On the contrary, due to the
lack of asymptotic freedom in QED, some of them are violated by the
Euler-Heisenberg Lagrangian at exponentially large magnetic field of
Planck scale, leading to appearance of ghosts, signifying the
instability of the magnetized vacuum. Superluminal excitations
(tachyons) might appear, too, but for the magnetic field exceeding
its instability threshold.
It is a surprise that the positive convexity property itself is not
violated at any value of the magnetic field. In Subsection C we
inspect two one-loop Lagrangians that are known to produce
spontaneous magnetic fields. One of them \cite{bms} relates to the
Yang-Mills theory taken against the uniform background formed by a
constant chromomagnetic field directed along a single isotopic
direction. The other \cite{kawati} is a one-loop Lagrangian of
electromagnetic field in interaction with a complex massless scalar
field taken in de Sitter space. We find that in the both cases the
spontaneous magnetization of the vacuum is due to the violation of
the positivity property of the Lagrangian convexity, prescribed by
the general principles of unitarity and causality. It is notable,
however, that in the Yang-Mills case the general properties of the
effective Lagrangian established in Section II other than the
convexity are well respected by the one-loop approximation, so
neither ghosts, nor tachyons appear. We associate this fact with the
asymptotic freedom of the underlying theory. In Subsection D another
Yang-Mills theory \cite{cabo}, \cite{cabshab} in a constant
homogeneous background is inspected, wherein
 the external field is this time supported by nonzero classical
sources and hence a special quantization procedure was used to
substitute for gauge invariance.

In the concluding Section IV we attempt a comparative discussion of
our approach with other  ways of introducing causality into
consideration.

\section{Generalities}
\subsection{Arbitrary dispersion $k_0\neq0$, $\textbf{k}\neq0$}
Let $\mathfrak{L}(z)$ be the nonlinear part of the effective
Lagrangian as a function of the two electromagnetic field invariants
$\mathfrak{F}$ and $\mathfrak{G}$ and, generally, of other Lorentz
scalars that can be formed by the electromagnetic field tensor
$F_{\mu\nu}$ and its space-time derivatives.  The total action  is
$S_{\rm tot}=\int L_{\rm tot}(z)\rmd^4z$, where $ L_{\rm
tot}(z)=-\mathfrak{F}(z)+ \mathfrak{L}(z).$ Once $-\mathfrak{F}$ is
the classical Lagrangian the correspondence principle implies that
\bee\label{corresp}\left.\frac{\delta
\Gamma}{\delta\mathfrak{F}}\right|_{\mathfrak{F}=\mathfrak{G}=0}=0,\eend
where $\Gamma=\int \mathfrak{L}(z)\rmd^4z$.

We consider the  background field, which is constant in time and
space and has only one nonvanishing invariant: $\mathfrak{F}\neq 0,
\mathfrak{G}=0$ (although $\mathfrak{G}$
 may be involved in  intermediate equations).
 This field is purely magnetic in
a special Lorentz frame, if $\mathfrak{F}>0$, and purely electric in
the opposite case, $\mathfrak{F}<0$. Such fields will be called
magnetic- or electric-like, respectively.

 Polarization operator is responsible for  small perturbations
 above the constant-field background. In accordance with the role
 of the effective action as the generating functional of vertex functions,
 the polarization operator  is defined as the
second variational derivative with respect to the vector potentials
$A_\mu$\bee\label{Pi} \Pi_{\mu\tau}(x,y)=
\left.\frac{\delta^2\Gamma}{\delta A_\mu(x)\delta A_\tau(y)}\right
|_{\mathfrak{G}=0, \mathfrak{F}= \rm const}.\eend The action
$\Gamma$ here is meant to be - prior to the two differentiations
over $A_\mu, A_\tau$ -  a functional containing field derivatives of
arbitrary order, but the fields are set constant after the
differentiations. Nevertheless, their derivatives do contribute into
the polarization operator (\ref{Pi}) leading to its complicated
dependence on the momentum $k$, the variable, Fourier conjugated to
$(x-y)$.

It follows from the translation- Lorentz-, gauge-, P- and
charge-invariance \cite{batalin, annphys,shabtrudy} that the Fourier
transform of the tensor (\ref{Pi}) is diagonal \bee\label{pimunu}
\Pi_{\mu\tau}(k,p)=\delta(k-p)\Pi_{\mu\tau}(k),\qquad
\Pi_{\mu\tau}(k)=\sum_{a=1}^3\varkappa_a(k)~\frac{\flat_\mu^{(a)}~
\flat_\tau^{(a)}}{(\flat^{(a)})^2}\eend  in the following
basis:\begin{eqnarray}\label{vectors} \flat^{(1)}_\mu=(F^2k)_\mu
k^2-k_\mu(kF^2k),\quad \flat_\mu^{(2)}=(\tilde{F}k)_\mu,\quad
\flat_\mu^{(3)}=(Fk)_\mu,\quad \flat_\mu^{(4)}=k_\mu,
\end{eqnarray} where $(\tilde{F}k)_\mu\equiv \tilde{F}_{\mu\tau}k_\tau$, $(Fk)_\mu\equiv
F_{\mu\tau}k_\tau$, $(F^2k)_\mu\equiv F^2_{\mu\tau}k_\tau$,
$kF^2k\equiv k_\mu F^2_{\mu\tau}k_\tau$, formed by the eigenvectors
of the polarization operator
\begin{eqnarray}\label{eigen}\Pi_{\mu\tau}~\flat^{(a)}_\tau=\varkappa_a(k)~\flat^{(a)}_\mu.
\end{eqnarray} We are working in Euclidian metrics with the results
analytically continued to Minkowsky space, hence we do not
distinguish between co- and contravariant indices. All eigenvectors
are mutually orthogonal, $\flat^{(a)}_\mu \flat_\mu^{(b)}\sim
\delta_{ab}$, this means that the first three ones are
4-transversal, $\flat^{(a)}_\mu k_\mu=0$; correspondingly
$\varkappa_4=0$ as a consequence of the 4-transversality of the
polarization operator. The unit matrix 
is decomposed as \bee\label{unit}
\delta_{\mu\tau}=\sum_{a=1}^4\frac{\flat_\mu^{(a)}~
\flat_\tau^{(a)}}{(\flat^{(a)})^2}\qquad {\rm or}\qquad
\delta_{\mu\tau}-\frac{k_\mu k_\tau}{k^2}
=\sum_{a=1}^3\frac{\flat_\mu^{(a)}~
\flat_\tau^{(a)}}{(\flat^{(a)})^2}.\eend The eigenvalues
$\varkappa_a(k)$ of the polarization operator are scalars and depend
on $\mathfrak{F}$ and on any two of the three momentum-containing
Lorentz invariants $k^2={\bf k}^2-k_0^2,\; kF^2k,\;k\tilde{F}^2k$ ,
subject to one relation $\frac{k\tilde{F}^2k}{2\mathfrak{F}}-k^2=
\frac{k{F}^2k}{2\mathfrak{F}}$.
 The squares of the eigenvectors
are\bee\label{squares}
(\flat^{(1)})^2=-k^2(kF^2k)((kF^2k)+2\mathfrak{F}k^2)=k^2k_\perp^2(2\mathfrak{F})^2(k_3^2-k^2_0),\nonumber\\
\quad (\flat^{(2)})^2=-(k\widetilde{F}^2k),\quad
(\flat^{(3)})^2=-(k{F}^2k)\quad\eend

The diagonal representation of the photon Green function as an exact
solution to the Schwinger-Dyson equation with the polarization
operator (\ref{pimunu}) taken for the kernel is (up to arbitrary
longitudinal part):
\begin{eqnarray}\label{photon2} D_{\mu\tau}(k)=
\sum_{a=1}^4 D_a(k)~\frac{\flat_\mu^{(a)}~
\flat_\tau^{(a)}}{(\flat^{(a)})^2},\nonumber\\
D_a(k)=\left\{\begin{tabular}{cc}$(k^2-\varkappa_a(k))^{-1},$&\qquad\;
$a=1,2,3$\\arbitrary, &~ $a$ = 4\end{tabular}\;.\right.
\end{eqnarray}
The dispersion equations that define the mass shells of the three
eigen-modes are
\bee\label{dispersion}\varkappa_a(k^2,\frac{k{F}^2k}{2\mathfrak{F}},\mathfrak{F})=k^2,\qquad
a=1,2,3. \eend

All the equations above are valid both for magnetic- and
electric-like cases, $\mathfrak{F}\lessgtr 0$, $\mathfrak{G}=0$. If,
specifically,  the magnetic-like background field $\mathfrak{F}> 0$,
$\mathfrak{G}=0$ is considered,  in the special frame the
field-containing invariants become \bee\label{special}\frac{k\tilde
{F}^2k}{2\mathfrak{F}}=k_3^2-k_0^2,\qquad\frac{k{F}^2k}{2\mathfrak{F}}=-k_\perp^2,\qquad
\mathfrak{F}=\frac{B^2}2,\eend where we directed the magnetic field
$\bf B$ along the axis 3, and the two-dimensional vector ${\bf
k}_\perp$ is the photon momentum projection onto the plane
orthogonal to it. On the contrary, if we deal with the electric-like
background field $\mathfrak{F}< 0$, $\mathfrak{G}=0,$ in the special
frame, where only electric field $\bf E$ exists and is
 directed along axis 3, we have, instead of (\ref{special}), the following
  relations for the background-field- and momentum-containing invariants
\bee\label{special2}\frac{k\tilde
{F}^2k}{2\mathfrak{F}}=k_\perp^2,\qquad\frac{k{F}^2k}{2\mathfrak{F}}=k_0^2-k_3^2,\qquad
\mathfrak{F}=\frac{-E^2}2,\eend where  the two-dimensional vector
${\bf k}_\perp$ now is the photon momentum projection onto the plane
orthogonal to $\bf E$. In the both cases the dispersion equations
(\ref{dispersion}) can be represented in the same form
\bee\label{dispersion2}\varkappa_a(k^2,k_\perp^2,\mathfrak{F})=k^2,\qquad
a=1,2,3 \eend and their solutions have the following general
structure, provided by relativistic invariance \bea
k_0^2=k_3^2+f_a(k_\perp^2),\hspace{5mm} a=1,2,3\label{law}. \eea

It is notable that the structure (\ref{law}) retains
\cite{newphysrev} when the second invariant is also nonzero,
$\mathfrak{G}\neq 0,$ this time the direction 3 being the common
direction of the background electric and magnetic fields in the
special reference frame, where these are mutually parallel.

The causal propagation requires that the modulus of the group
velocity, calculated on each mass shell (\ref{law}), be less or
equal to the speed of light in the free vacuum $c=1$:
\bee\label{group}|\textbf{v}_{\rm gr}|^2=\left(\frac{\partial
k_0}{\partial k_3}\right)^2+\left|\frac{\partial k_0}{\partial
\textbf{k}_\perp}\right|^2=\frac{k_3^2}{k_0^2}+\left|\frac{\textbf{k}_\perp}{k_0}\cdot
f_a^\prime\right|^2
=\frac{k_3^2+k_\perp^2\cdot(f_a^\prime)^2}{k_3^2+f_a(k_\perp^2)}\leq
1,\eend where $f_a^\prime=\rmd f_a(k_\perp^2)/\rmd k_\perp^2$. This
imposes the obligatory condition on the form and location of the
dispersion curves (\ref{law}), i.e. on the function
$f_a(k_\perp^2)$, to be fulfilled within every reasonable
approximation (remind that $k_3^2+f_a(k_\perp^2)\geq 0 $ due to
(\ref{law})) :\bee\label{causality} k_\perp^2\left(\frac{\rmd
f_a(k_\perp^2)}{\rmd k_\perp^2}\right)^2\leq f_a(k_\perp^2).\eend
The admissible disposition of dispersion curves was considered by us
for the general case of $\mathfrak{G}\neq0$ in detail in
\cite{newphysrev}. We found that the massless branches of these
curves ("photons"), whose existence is always guarantied by the
gauge invariance, for every polarization mode are outside  the light
cone (or on it) in the momentum space, $k^2=0$, whereas the massive
branches all should pass below a certain curve in the plane
$(k_0^2-k_3^2, k_\perp^2)$, where $k_3$ and $\bf k_\perp$ are the
excitation momentum components along and across the direction of the
background magnetic and electric fields in the special frame, where
these are mutually parallel. We also discussed in that reference why
and to what extent the restriction on the group velocity may be
equivalent to causality.

Now we proceed by imposing the condition, to be referred to, as
unitarity, that the residues of the photon propagator
(\ref{photon2}) in the poles corresponding to every photon mass
shell (\ref{dispersion}) be nonnegative - the positive definiteness
of the norm of every elementary excitation of the vacuum. This
requirement implies:\bee\label{unitarity}
1-\left.\frac{\partial\varkappa_a(k^2,k_\perp^2,\mathfrak{F})}{\partial
k^2}\right|_{k_0^2-k_3^2=f_a(k_\perp^2)}\geq 0.\eend In the next
subsection we shall consider the consequences of requirements
(\ref{causality}) and (\ref{unitarity}) as these manifest themselves
in the properties of the effective Lagrangian in the infrared limit.

\subsection{Infrared limit: properties of the Lagrangian as a
function of constant fields} Hitherto, we were dealing with the
elementary excitation of arbitrary 4-momentum $k_\mu.$ To get the
(infrared) behavior of the polarization operator at $k_\mu\sim 0$ it
 is sufficient to have at one's disposal the
effective Lagrangian as a function of constant field strengthes,
since their space- and time-derivatives, if included in the
Lagrangian, would supply extra powers of the momentum $k$ in the
expression (\ref{Pi}) for the polarization operator.  Our goal is to
establish some inequalities imposed on the derivatives of the
effective Lagrangian $\mathfrak{L}$ over the constant fields by the
requirement (\ref{causality}) that any elementary excitation of the
vacuum should not propagate with the group velocity larger than
unity and the requirement (\ref{unitarity}) that the residue of the
Green function be positive in the photon pole. To proceed beyond
this limit we had to include the space and time derivatives of the
fields into the Lagrangian. Then, utilizing the same  requirements
(\ref{causality}), (\ref{unitarity}) the results concerning the
convexity of the effective Lagrangian with respect to the constant
fields to be obtained below, might be, perhaps, extended to
convexities with respect to the derivative-containing field
variables.

Aiming at the infrared limit we do not include time- and
space-derivatives of the field strengthes in the equations that
follow. Using the definition $F_{\alpha\beta}(z)=\partial_\alpha
A_\beta(z)-\partial_\beta A_\alpha(z)$ we find
\bee\label{fi}\frac\delta {\delta A_\mu(x)}\int
\mathfrak{F}(z)\rmd^4z=\int F_{\alpha\mu}(z)\frac{\partial}{\partial
z_\alpha}\delta^4(x-z)\rmd^4z,\nonumber\\ \frac\delta {\delta
A_\mu(x)}\int \mathfrak{G}(z)\rmd^4z=\int
\tilde{F}_{\alpha\mu}(z)\frac{\partial}{\partial
z_\alpha}\delta^4(x-z)\rmd^4z.\eend Then, for the first variational
derivative of the action one has\bee\label{firstder}\frac{\delta
\Gamma} {\delta A_\mu(x)}=
\int\left[\frac{\partial\mathfrak{L}(\mathfrak{F}(z),\mathfrak{G}(z))}{\partial
\mathfrak{F}(z)}F_{\alpha\mu}(z)+\frac{\partial\mathfrak{L}(\mathfrak{F}(z),\mathfrak{G}(z))}{\partial
\mathfrak{G}(z)}\tilde{F}_{\alpha\mu}(z)\right]\frac{\partial}{\partial
z_\alpha}\delta^4(x-z)\rmd^4z.\eend By repeatedly applying eq.
(\ref{firstder})  we get for the infrared (IR) limit of the
polarization operator in a constant external field
\bee\label{secdir}\Pi^{\rm
IR}_{\mu\tau}(x,y)=\left.\frac{\delta^2\Gamma}{\delta A_\mu(x)\delta
A_\tau(y)}\right |_{\mathfrak {F},\mathfrak {G} =\rm const}=
\left\{\frac{\partial\mathfrak{L}(\mathfrak{F}(z),\mathfrak{G}(z))}{\partial
\mathfrak{F}(z)}\left(\frac{\partial^2}{\partial x_\tau\partial
x_\mu}-\Box\delta_{\mu\tau}\right)\right.
-\nonumber\\\nonumber\\-\left.\frac{\partial^2\mathfrak{L}(\mathfrak{F}(z),\mathfrak{G}(z))}{\partial
(\mathfrak{F}(z))^2}\left(F_{\alpha\mu}\frac{\partial }{\partial
x_\alpha}\right)\left(F_{\beta\tau}\frac{\partial }{\partial
x_\beta}\right)-
\frac{\partial^2\mathfrak{L}(\mathfrak{F}(z),\mathfrak{G}(z))}{\partial
(\mathfrak{G}(z))^2}\left(\tilde{F}_{\alpha\mu}\frac{\partial
}{\partial x_\alpha}\right)\left(\tilde{F}_{\beta\tau}\frac{\partial
}{\partial x_\beta}\right)\right.
-\nonumber\\
\nonumber\\
-\left.\frac{\partial^2\mathfrak{L}(\mathfrak{F}(z),\mathfrak{G}(z))}{\partial
\mathfrak{F}(z)\partial \mathfrak{G}(z)}\left[
\left({F}_{\alpha\mu}\frac{\partial }{\partial
x_\alpha}\right)\left(\tilde{F}_{\beta\tau}\frac{\partial }{\partial
x_\beta}\right)+\left(\tilde{F}_{\alpha\mu}\frac{\partial }{\partial
x_\alpha}\right)\left({F}_{\beta\tau}\frac{\partial }{\partial
x_\beta}\right) \right]\right\}_{F=\rm
const}\delta^4(x-y).\quad\eend The P-invariance requires that the
effective Lagrangian should be an even function of the pseudoscalar
$\mathfrak{G}$. Hence the contribution of the last term in  eq.
(\ref{secdir}) -- the one in front of the square bracket -- vanishes
for the "single-invariant" fields with $\mathfrak{G}=0$ under
consideration.

Thus, we find for the infrared limit of the polarization operator in
the magnetic- or electric-like field in the momentum representation,
$\Pi^{\rm IR}_{\mu\tau}(k,p)=\delta(k-p)\Pi^{\rm
IR}_{\mu\tau}(k),$\bee \label{fourier} \Pi^{\rm IR}_{\mu\tau}(k)=
\left(\frac{\rmd\mathfrak{L}(\mathfrak{F},0)}{\rmd
\mathfrak{F}}(\delta_{\mu\tau}k^2-k_\mu k_\tau )
+\frac{\rmd^2\mathfrak{L}(\mathfrak{F},0)}{\rmd\mathfrak{F}^2}(F_{\mu\alpha}k_\alpha)(F_{\tau\beta}k_\beta)\right.
+\nonumber\\\nonumber\\+\left.\left.\frac{\partial^2\mathfrak{L}(\mathfrak{F},
\mathfrak{G})}{\partial\mathfrak{G}^2}\right|_{\mathfrak{G}=0}
(\tilde{F}_{\mu\alpha}k_\alpha)(\tilde{F}_{\tau\beta}k_\beta)\right).\eend
Here the scalar $\mathfrak{F}$ and the tensors $F, \tilde{F}$ are
already set to be  space- and time-independent. By comparing this
with (\ref{pimunu}) we identify the eigenvalues of the polarization
operator in the infrared limit as
\bee\label{kappa}\left.\varkappa_1(k^2,kF^2k,\mathfrak{F})\right|_{k\rightarrow
0}= k^2\frac{\rmd\mathfrak{L}(\mathfrak{F},0)}{\rmd\mathfrak{F}},
\nonumber\\\nonumber\\\left.\varkappa_2(k^2,kF^2k,\mathfrak{F})\right|_{k\rightarrow
0} = k^2\frac{\rmd\mathfrak{L}(\mathfrak{F},0))}{\rmd\mathfrak{F}}-
(k\tilde{F}^2k)\left.\frac{\partial^2\mathfrak{L}(\mathfrak{F},\mathfrak{G})}
{\partial\mathfrak{G}^2}\right|_{\mathfrak{G}=0},\nonumber\\\nonumber\\
\left.\varkappa_3(k^2,kF^2k,\mathfrak{F})\right|_{k\rightarrow 0}=
k^2\frac{\rmd\mathfrak{L}(\mathfrak{F},0)}{\rmd\mathfrak{F}}-
(kF^2k)\frac{\rmd^2\mathfrak{L}(\mathfrak{F},0)}
{\rmd\mathfrak{F}^2}.\eend This is the leading behavior  of the
polarization operator in the magnetic-like field near zero-momentum
point $k_\mu =0$. Every eigenvalue $\varkappa_a$ is a linear
function of $k_\perp^2$ and of $k_0^2-k_3^2$ , hence
$\varkappa_a(0,0,\mathfrak{F})=0$ for every $a=1,2,3$. This is a
nondispersive approximation, since the refraction index (squared)
$n^2_a$  defined for photons of each mode \emph{a} on the mass shell
(\ref{law}) as \bea\label{refrindex} n_a^2\equiv\frac{|{\bf
k}|^2}{k_0^2}
=1+\frac{k_\bot^2-f_a(k_\bot^2)}{k_0^2} \eend is frequency- and
momentum-independent in the infrared limit under consideration.

For the sake of completeness, we give the same eqs. (\ref{kappa})
also in terms of the invariant variables
 \bee\label{variables}\mathcal{B}=
\sqrt{\mathfrak{F}+\sqrt{\mathfrak{F}^2+\mathfrak{G}^2}}\qquad
\mathcal{E}=\sqrt{-\mathfrak{F}+\sqrt{\mathfrak{F}^2+\mathfrak{G}^2}}\eend
that are, respectively,  the  magnetic and electric fields in the
Lorentz frame, where these are parallel. Then, with the notation
${\mathfrak{\widetilde{L}}}(\mathcal{B},\mathcal{E})=\mathfrak{L}(\mathfrak{F},\mathfrak{G})$
the coefficients in (\ref{kappa}) are :\bee\label{subst}
 \frac{\rmd\mathfrak{L}(\mathfrak{F},0)}{\rmd\mathfrak{F}}=
\frac{1}{\mathcal{B}}\frac{\rmd \mathfrak{\widetilde{L}}(\mathcal{B},0)}{\rmd \mathcal{B}},\nonumber\\
 \frac{\rmd^2\mathfrak{L}(\mathfrak{F},0)}{\rmd\mathfrak{F}^2}=
 \frac 1{2\mathfrak{F}}\left(\frac{\rmd^2\mathfrak{\widetilde{L}}(\mathcal{B},0)}{\rmd \mathcal{B}^2}-
 \frac{~~\rmd \mathfrak{\widetilde{L}}(\mathcal{B},0)}{\mathcal{B}\rmd \mathcal{B}}\right),\nonumber\\
 \left.\frac{\partial^2\mathfrak{L}(\mathfrak{F},
\mathfrak{G})}{\partial\mathfrak{G}^2}\right|_{\mathfrak{G}=0}=
\frac{1}{2\mathfrak{F}}\left. \left(\frac
1{\mathcal{E}}\frac{\partial
\mathfrak{\widetilde{L}}(\mathcal{B},\mathcal{E})}{\partial
\mathcal{E}}\right)\right|_{\mathcal{E}=0}+
 \frac{1}{2\mathfrak{F}}\frac 1{\mathcal{B}}\frac{~\rmd \mathfrak{\widetilde{L}}(\mathcal{B},0)}{\rmd \mathcal{B}}.\eend

 At this step we turn  to the special case of magnetic-like background
 and shall be sticking to it until the end of the present Subsection,
 keeping the extension of some results to the electric-like case
$\mathfrak{F}<0$  to the next Subsection C.

The dispersion curves $f_a(k_\perp^2)$ near the origin may be found
by solving equations (\ref{dispersion}) in the special frame with
the right-hand sides taken as (\ref{kappa}) and with eqs.
(\ref{special}) taken into account. This gives the linear functions
for  photons of modes 2 and 3
\bee\label{linear2}f_2(k_\perp^2)=k_\perp^2\left(\frac{1-\mathfrak{L_F}}
{1-\mathfrak{L_{F}}+2\mathfrak{F}\mathfrak{L_{GG}}}\right), \eend
\bee\label{linear3}f_3(k_\perp^2)=k_\perp^2\left(1-\frac{2\mathfrak{F}\;
\mathfrak{L_{FF}}}{1-\mathfrak{L_{F}}}\right), \eend where we are
using the notations
$\mathfrak{L_{FF}}=\frac{\rmd^2\mathfrak{L}(\mathfrak{F},0)}
{\rmd\mathfrak{F}^2},\quad$
$\mathfrak{L_{F}}=\frac{\rmd\mathfrak{L}(\mathfrak{F},0))}{\rmd\mathfrak{F}},\quad
\mathfrak{L_{GG}}=\left.\frac{\partial^2\mathfrak{L}(\mathfrak{F},\mathfrak{G})}
{\partial\mathfrak{G}^2}\right|_{\mathfrak{G}=0}.$ As for mode 1,
the dispersion equation in the present approximation has only the
trivial solution $k^2=0$ that makes the vector potential
$\flat_\mu^{(1)}$ corresponding to it purely longitudinal, with no
electromagnetic field  carried by the mode. This is a nonpropagating
mode in the infrared limit (it is also nonpropagating within the
one-loop approximation beyond this limit; however,
massive-positronium  solutions in mode 1 do propagate \cite{ass}).

The unitarity condition (\ref{unitarity}), as applied to mode 2,
gives via the second equation in (\ref{kappa})
\bee\label{unitarity2}1-\mathfrak{L_F}+2\mathfrak{F}\mathfrak{L_{GG}}\geq
0. \eend 
Then,  from the  behavior of the dispersion curve (\ref{linear2})
and
the causality (\ref{causality}) 
it follows that \bee\label{unicaus}1-\mathfrak{L_F}\geq 0 \eend
and \bee\label{min2}\mathfrak{L_{GG}} \geq 0.\eend 
(Remind that for the magnetic-like  case under consideration one has
$\mathfrak{F}>0$.) 

Analogously, the unitarity condition (\ref{unitarity}), as applied
to mode 3, gives via the third equation in (\ref{kappa}) again the
result (\ref{unicaus}). (This inequality also provides the
positiveness of the norm of the non-propagating mode 1.) Then from
the  behavior of the dispersion curve (\ref{linear3}) and the
causality (\ref{causality}) it follows that \bee\label{unitarity3}
1-\mathfrak{L_F}-2\mathfrak{F}\mathfrak{L_{FF}}\geq 0\eend and
\bee\label{LGG}\mathfrak{L_{FF}} \geq 0.\eend

Inequalities eq.(\ref{unicaus}),  eq.(\ref{unitarity3}) together
provide that all the three residues of the photon Green function in
the complex plane of $k_\perp^2$, the same as in the complex plane
of $(k_3^2-k_0^2)$, eq.(\ref{unitarity}), are also nonnegative
\bee\label{residue2}
1-\left.\frac{\partial\varkappa_a(k^2,k_\perp^2,\mathfrak{F})}{\partial
k_\perp^2}\right|_{k_0^2-k_3^2=f_a(k_\perp^2)}\geq 0,\eend at least
in the infrared limit. We do not know whether this statement is
prescribed by general principles and therefore might be expected to
hold beyond this limit.

 Relations (\ref{min2}),  (\ref{LGG}) indicate that
the Lagrangian is a positively (downward) convex function of
$\mathfrak{F}$ for any $\mathfrak{F} > 0$ and of $\mathfrak{G}$ in
the point $\mathfrak{G}=0$.

Relations (\ref{unitarity2}), (\ref{unicaus}), (\ref{unitarity3})
indicate positiveness of various dielectric and magnetic
permittivity constants that control electro- and magneto-statics of
charges and currents of certain configurations. Eqs. (\ref{kappa})
imply that the quantities that are subject to the inequalities
(\ref{unitarity2}), (\ref{unicaus}) and (\ref{unitarity3}) are
expressed in terms of  different infra-red limits of the
polarization operator eigenvalues as \bee\label{from17a}
1-\mathfrak{L_F}=\lim_{k_\perp^2\rightarrow0}
\left(1-\frac{\left.\varkappa_2\right|_{k_0=k_3=0}}{k_\perp^2}\right)\equiv\varepsilon_{\rm
tr}(0),\nonumber\\
1-\mathfrak{L_F}=\lim_{k_\perp^2\rightarrow0}
\left(1-\frac{\left.\varkappa_1\right|_{k_0=k_3=0}}{k_\perp^2}\right)\equiv\left(\mu^{\rm
w}_{\rm tr}(0))\right)^{-1},\nonumber\\
1-\mathfrak{L_F}=\lim_{k_3^2\rightarrow 0}
\left(1-\frac{\left.\varkappa_3\right|_{k_0=k_\perp=0}}{k_3^2}\right)\equiv\left(\mu^{\rm
pl}_{\rm long}(0)\right)^{-1}, \eend
 \bee\label{from17b}
1-\mathfrak{L_F}+2\mathfrak{F}\mathfrak{L_{GG}}=\lim_{k_3^2\rightarrow0}
\left(1-\frac{\left.\varkappa_2\right|_{k_0=k_\perp=0}}{k_3^2}\right)\equiv\varepsilon_{\rm
long}(0),\eend
\bee\label{from17c}
1-\mathfrak{L_F}-2\mathfrak{F}\mathfrak{L_{FF}}=\lim_{k_\perp^2\rightarrow0}
\left(1-\frac{\left.\varkappa_3\right|_{k_0=k_3=0}}{k_\perp^2}\right)\equiv\left(\mu^{\rm
pl} _{\rm tr}(0)\right)^{-1}.\eend It is demonstrated in  Appendix
of Ref. \cite{convexity}  that $\varepsilon_{\rm long}$ and
$\varepsilon_{\rm tr}$ are dielectric constants responsible for
polarizing the homogeneous electric fields parallel and orthogonal
to the external magnetic field, which are produced, respectively, by
uniformly charged planes (sufficiently far from them as compared
with the formation length of the polarization operator), oriented
across the external magnetic field and parallel to it, see eqs.(123)
and (125) of \cite{convexity}. These are determined by the
eigenvalue $\varkappa_2$, the virtual photons of the mode 2 being
carriers of electrostatic force.

The quantity $\mu^{\rm w}_{\rm tr}(0)$ is the magnetic permittivity
constant responsible for attenuation of the magnetic field produced
by a constant current concentrated on a line, parallel to the
external magnetic field, sufficiently far from the current-carrying
line, see Ref. \cite{convexity} eq.(110) with  $\mu(0)$ replaced by
 $\mu^{\rm w}_{\rm tr} (0)$ in it. The same quantity $\mu^{\rm w}_{\rm tr}(0)$
 governs the constant magnetic field of a plane current flowing along the external field.
 This magnetic permittivity  is determined by the mode 1.
 The other two magnetic permittivities, $\mu^{\rm
pl} _{\rm long}(0)$ and $\mu^{\rm pl} _{\rm tr}(0)$ are determined
by the mode 3. The permittivity $\mu^{\rm pl} _{\rm tr}(0)$ is
responsible for  remote attenuation of the  magnetic field produced
by a constant current, homogeneously concentrated on a plane,
parallel to the external magnetic field, and flowing in the
direction transverse to it, see Ref. \cite{convexity} eq.(135). This
magnetic field is homogeneous and parallel to the external field.
Finally, permittivity $\mu^{\rm pl} _{\rm long}(0)$ is responsible
for remote attenuation of the  magnetic field produced by a constant
straight current, homogeneously concentrated on a plane, transverse
to the external magnetic field, see Ref. \cite{convexity} eq.(138).
This field is also homogeneous. Virtual photons of the modes 1 and 3
are carriers of magneto-static force.

By using the  wordings "sufficiently far" and "remote" we mean
distances from the corresponding sources that essentially exceed a
characteristic length of an underlying microscopic theory, wherein
the linear response is formed. In a material medium that may be an
interatomic distance; in perturbative QED this is the electron
Compton length.

Relations (\ref{from17a}), (\ref{from17b}), (\ref{from17c}) mean
that the inequalities (\ref{unitarity2}), (\ref{unicaus}) and
(\ref{unitarity3}) signify the positiveness of all the
characteristic  permittivities of the magnetized vacuum, which was
derived above on general basis. Besides, thanks to (\ref{from17a}),
there exists the equality between one dielectric and two (inverse)
magnetic permittivities\bee\label{equality}\varepsilon_{\rm
tr}(0)=\left(\mu^{\rm w}_{\rm tr}(0)\right)^{-1}=\left(\mu^{\rm
pl}_{\rm long}(0)\right)^{-1}.\eend The first equality here is a
direct consequence of the invariance under the Lorentz boost along
the magnetic field in the special frame (see eq. (73) in
\cite{convexity} and can be extended to the permittivity functions
as defined in \cite{convexity}  by (128) and the right equation
(121), $\varepsilon_{\rm tr}(k_\perp^2)=\left(\mu^{\rm w}_{\rm
tr}(k_\perp^2)\right)^{-1}$.

Relations (\ref{from17a}) -- (\ref{from17c}) together with
(\ref{min2}), (\ref{unitarity3}) also mean that the longitudinal
dielectric constant should be always larger than the transversal one
\bee\label{epsil}\varepsilon_{\rm long}(0)\geq \varepsilon_{\rm
tr}(0),\eend while the magnetic permittivities should satisfy the
opposite inequality\bee\label{mu}\mu^{\rm pl}_{\rm tr}(0)\geq
\mu^{\rm pl}_{\rm long}(0).\eend
\subsection{Electriclike background field}
In this subsection we shall see how the inequalities
(\ref{unitarity2})--(\ref{LGG})  derived in the previous Subsection
are extended to the negative domain of the invariant $\mathfrak{F}$.

  Bearing in mind eqs. (\ref{special2}) we may solve again  dispersion equations (\ref{dispersion2})
   using eqs. (\ref{kappa}) to get  the photon dispersion curves in the electriclike background field in the
infrared approximation. For mode 2 this results in
\bee\label{disp2}k_0^2-k_3^2=k_\perp^2\left(1+\frac{2\mathfrak{FL_{GG}}}{1-\mathfrak{L_F}}\right),\eend
while  for mode 3 in
\bee\label{disp3}k_0^2-k_3^2=k_\perp^2\left(\frac{1-\mathfrak{L_F}}{1-\mathfrak{L_F}-2\mathfrak{FL_{FF}}}\right)\eend
(compare this with (\ref{linear2}), (\ref{linear3})). The unitarity
relation (\ref{unitarity}) applied to mode 2 leads to the inequality
(\ref{unicaus}). The causality condition (\ref{causality}), when
applied to (\ref{disp2}) requires that\bee\label{ineq}
\left(1+\frac{2\mathfrak{FL_{GG}}}{1-\mathfrak{L_F}}\right)^2\leq
\left(1+\frac{2\mathfrak{FL_{GG}}}{1-\mathfrak{L_F}}\right).\eend
This implies that the right-hand side of the inequality (\ref{ineq})
be positive and  thus the both sides can be divided on it. Then the
inequality (\ref{ineq}) becomes the inequality
(\ref{unitarity2})\bee
\left(1+\frac{2\mathfrak{FL_{GG}}}{1-\mathfrak{L_F}}\right)<1.\eend
In view of (\ref{unicaus}) this means that $2\mathfrak{FL_{GG}}<0.$
Once $\mathfrak{F}$ is negative for the electric -like case under
consideration now, we come again to the convexity condition
(\ref{min2}), now in the domain of negative $\mathfrak{F}$. By
applying the same procedure to mode 3 we quite analogously reproduce
eqs. (\ref{unitarity3}) and (\ref{LGG}).
\subsection{Energy-momentum conditions}
Now we proceed with describing general restrictions imposed by the
physical requirement that the energy density of elementary
excitations of the magnetic-like  background (magnetized vacuum) be
nonnegative ("weak energy condition" in terms of Ref.
\cite{hawking})\bee\label{endensity}t_{00}\geq 0\eend and that their
energy-momentum flux density be non-spacelike ("dominant energy
condition" of Ref.
\cite{hawking}))\bee\label{poynting}t_{0\nu}^2\leq 0\eend in order
to compare the results with the conclusions of Subsection B.

We have to define the energy-momentum tensor $t_{\mu\nu}(x)$ of
small perturbations of the background field by first defining their
 Lagrangian. The total effective Lagrangian $L_{\rm
tot}=-\mathfrak{F}+\mathfrak{L}$
 expanded near the background constant magnetic field contributes
 into the total action -- in view
  of the definition (\ref{Pi}) -- the following
 correction, quadratic in the small perturbation $a_\mu(x)$ above the
 background:  \bee\label{qudrcor}S_{\rm tot}^{\rm
 sqr}=\frac 1{2}\int a_\mu(x)\{-\left(\delta_{\mu\nu}\partial^2_\alpha-\frac\partial{\partial x_\mu}
\frac\partial{\partial
y_\nu}\right)\delta(x-y)+\Pi_{\mu\nu}(x,y)\}a_\nu(y)\rmd^4x\rmd^4y.\eend
The field intensity of the perturbation will be denoted as
$f_{\mu\nu}=\partial_\mu a_\nu - \partial_\nu a_\mu$. Using the
diagonal form of the polarization operator (\ref{pimunu}) we get in
the momentum representation \bee\label{nonlocal}L_{\rm tot}^{\rm
sqr}(k)=\frac 1{4}f^2+\frac1{4}\left(-\frac{\varkappa_1}{k^2}f^2
+\frac{\varkappa_1-\varkappa_2}{2k\widetilde{F}^2k}((f\widetilde{F}))^2+
\frac{\varkappa_1-\varkappa_3}{2k{F}^2k}((fF))^2\right).\eend Here
the notations are used:
$(fF)_{\mu\nu}=f_{\mu\alpha}F_{\alpha\nu}=(Ff)_{\nu\mu},~$$
(fF)=(fF)_{\mu\mu}=(Ff)$,
$~f^2_{~\mu\nu}=f_{\mu\alpha}f_{\alpha\nu},~$
$f^2=f^2_{~\mu\mu}=-(f_{\mu\nu})^2$, and we have exploited the
relations $f^2=-2a_\mu(k^2\delta_{\mu\nu}-k_\mu k_\nu)a_\nu$,
$(fF)=2(aFk)$. This Lagrangian is nonlocal, since it depends on
momenta in a complicated way, in other words, it depends highly
nonlinearly on the derivatives with respect to coordinates. It
becomes local if we restrict ourselves to the infrared limit by
substituting eqs.(\ref{kappa}) into it. Then the quadratic
Lagrangian acquires the very compact form\bee\label{compact}L_{\rm
tot}^{\rm sqr}=\frac 1{4}f^2(1-\mathfrak{L_F})+\frac {1}{8}\left(
\mathfrak{L_{GG}}((f\widetilde{F}))^2+
\mathfrak{L_{FF}}((fF))^2\right).\eend This Lagrangian, quadratic in
the field $f_{\mu\nu}(x)$, does not contain its derivatives,
$F_{\mu\nu},\widetilde{F}_{\mu\nu},\mathfrak{L_F},\mathfrak{L_{GG}}$
and $\mathfrak{L_{FF}}$ being constants depending upon the
background field alone. It governs small-amplitude low-frequency and
low-momentum perturbations of the magnetized vacuum, free of or
created by
 small sources. It might be obtained also directly by
calculating the second derivative (\ref{Pi}) of the Lagrangian
defined on constant fields \cite{dipiazza}.

Once the  background is translation-invariant, there is a conserved
energy-momentum tensor $t_{\mu\nu}(x)$ of the field $f_{\mu\nu}$
provided by the  Noether theorem by considering variations of this
field. Applying the  standard definition  of the energy-momentum
tensor to the field of small perturbation $a_\mu$  and to its
Lagrangian (\ref{compact}) we get \bee\label{noether1}
t_{\mu\nu}(x)= -\frac{\partial L_{\rm tot}^{\rm sqr}}{\partial
(\partial a_\alpha/\partial x_\nu)}\frac{\partial a_\alpha}{\partial
x_\mu}+\delta_{\mu\nu}L_{\rm tot}^{\rm sqr}=\nonumber\\
=-\frac{\partial a_\alpha}{\partial
x_\mu}\left(f_{\alpha\nu}(1-\mathfrak
{L_F})+\frac1{2}(f\widetilde{F})\mathfrak{L_{GG}}\widetilde{F}_{\alpha\nu}
+\frac1{2}(f{F})\mathfrak{L_{FF}}{F}_{\alpha\nu}\right)
+\delta_{\mu\nu}L_{\rm tot}^{\rm sqr}.\eend The Maxwell equations
for small sourceless perturbations of the magnetized vacuum are
\bee\label{maxwell}\frac{\delta L_{\rm tot}^{\rm sqr}}{\delta
a_\alpha}=\frac{\partial}{\partial x_\nu}\frac{\partial L_{\rm
tot}^{\rm sqr}}{\partial (\partial a_\alpha/\partial
x_\nu)}=\frac{-\partial}{\partial
x_\nu}\left(f_{\alpha\nu}(1-\mathfrak
{L_F})+\frac1{2}(f\widetilde{F})\mathfrak{L_{GG}}\widetilde{F}_{\alpha\nu}
+\frac1{2}(f{F})\mathfrak{L_{FF}}{F}_{\alpha\nu}\right)=0.\quad\eend
We are going to use the standard indeterminacy in the definition of
the energy-momentum tensor to let it depend only on the field
strength $f_{\mu\nu},$ and not on its potential. To this end we  add
the quantity (the designation $\doteq$ below means "equal up to full
derivative") \bee\label{quantity}\frac{\partial L_{\rm tot}^{\rm
sqr}}{\partial (\partial a_\alpha/\partial x_\nu)}\frac{\partial
a_\mu}{\partial x_\alpha}\doteq-a_\mu\frac\partial{\partial
x_\alpha}\frac{\partial L_{\rm tot}^{\rm sqr}}{\partial (\partial
a_\alpha/\partial x_\nu)}=\nonumber\\=a_\mu\frac\partial{\partial
x_\alpha} \{f_{\alpha\nu}(1-\mathfrak
{L_F})+\frac1{2}(f\widetilde{F})\mathfrak{L_{GG}}\widetilde{F}_{\alpha\nu}
+ \frac1{2}(f{F})\mathfrak{L_{FF}}{F}_{\alpha\nu}\}\eend to
(\ref{noether1}), that disappears due to the Maxwell equations
(\ref{maxwell}), taking into account the antisymmetricity of the
expression inside the braces. Hence the energy-momentum tensor may
be equivalently written as\bee\label{en-mom} t_{\mu\nu}(x)=
-f^2_{~\mu\nu}(1-\mathfrak
{L_F})-\frac1{2}(f\widetilde{F})\mathfrak{L_{GG}}(f\widetilde{F})_{\mu\nu}
-\frac1{2}(f{F})\mathfrak{L_{FF}}(f{F})_{\mu\nu}+\nonumber\\
+\frac{\delta_{\mu\nu}}4\left( f^2(1-\mathfrak{L_F})+
\frac1{2}\mathfrak{L_{GG}}((f\widetilde{F}))^2+
\frac1{2}\mathfrak{L_{FF}}((fF))^2\right).\qquad\qquad\eend This
tensor is traceless, $t_{\mu\mu}=0$. It obeys the continuity
equation with respect to the {\em second~}
index\bee\label{contin}\frac{\partial t_{\mu\nu}}{\partial
x_\nu}=0\eend owing to the Maxwell equations (\ref{maxwell}). Hence,
the 4-momentum vector obtained by integrating $t_{0\mu}$ over the
spatial volume $\rmd^3x$ conserves in time.

Let us take (\ref{en-mom}), first, on the monochromatic -- with
4-momentum $k_\mu$ -- real solution of the Maxwell equations
(\ref{maxwell}) that belongs to the eigen-mode 3:
$f_{\mu\nu}^{(3)}=k_\mu\flat^{(3)}_\nu-k_\nu\flat^{(3)}_\mu$. One
has
$(f^{(3)}{F})_{\mu\nu}=\flat^{(3)}_\mu\flat^{(3)}_\nu-k_\mu(F^2k)_\nu$,
$(f^{(3)}{F})=-2(kF^2k)$,
$(f^{(3)})^2_{\;\mu\nu}=-k^2\flat^{(3)}_\mu\flat^{(3)}_\nu+k_\mu
k_\nu(kF^2k),$ $(f^{(3)})^2=2k^2(kF^2k)$,
$(f^{(3)}\widetilde{F})=0$. With the substitution
$f_{\mu\nu}=f_{\mu\nu}^{(3)}$ the Maxwell equation (\ref{maxwell})
is satisfied, when \bee\label{sat}\flat^{(3)}_\alpha\{ k^2
(1-\mathfrak {L_F})+(kF^2k)\mathfrak{L_{FF}}\}=0,\eend i.e.,
naturally, on the dispersion curve (\ref{linear3}) for mode 3. 
It is seen that the Lagrangian (\ref{compact}) disappears on the
mass shell of mode 3, $L_{\rm tot}^{{\rm sqr} (3)}=0$. Then, the
reduction of the energy momentum tensor (\ref{en-mom}) onto this
mode, $t^{(3)}_{\mu\nu}(x),$ should be written with its
$\delta_{\mu\nu}$ part
 dropped: \bee\label{T3}t^{(3)}_{\mu\nu}(x)= (1-\mathfrak
{L_F})(k^2\flat^{(3)}_\mu\flat^{(3)}_\nu-k_\mu k_\nu(kF^2k))
+(kF^2k)\mathfrak{L_{FF}}(\flat^{(3)}_\mu\flat^{(3)}_\nu-k_\mu(F^2k)_\nu)
.\eend Then, after omitting the  common factor $-(kF^2k)$ equal to
$2\mathfrak{F}k_\perp^2>0$ in a magnetic field, and to
$2\mathfrak{F}(k_0^2-k_3^2)>0$ in an electric field, and using the
mass shell equation once again, we get \bee\label{T0mu}
t^{(3)}_{\mu\nu }(x)= (1-\mathfrak {L_F})k_\mu k_\nu +
k_\mu\mathfrak{L_{FF}}(F^2k)_\nu. \eend Although we referred to the
magnetic-like background above in
 this Subsection, all the equations written in it up
to now remain, as a matter of fact,  valid also for the
electric-like case. In the rest of this Subsection we actually
specialize to the magnetized vacuum, although the conclusions may be
readily extended to cover  the electrified vacuum, as well.
  When $\mathfrak{F}>0,$ in the special frame $(F^2k)_{0,3}=0,$
$(F^2k)_{1,2}=-2\mathfrak{F}k_{1,2}.$
   It is convenient to write the energy-momentum density vector in
components (counted as 0,1,2,3 downwards)
\bee\label{components}t_{0\nu}^{(3)}=k_0\left(\begin{tabular}{c}$k_0(1-\mathfrak{L_F})
$\\$k_1(1-\mathfrak{L_F}-2\mathfrak{FL_{FF})}$\\$k_2(1-\mathfrak{L_F}-2\mathfrak{FL_{FF})}$
\\$k_3(1-\mathfrak{L_F})$\end{tabular}\right)_\nu.\eend
The spacial part of this vector density is parallel to the group
velocity ${\bf v}_{\rm gr}^{(3)}= (\rmd k_0/\rmd{\bf k})$ calculated
on the mode-3-mass-shell as defined by the dispersion law
(\ref{law}), (\ref{linear3})\bee\label{parallel}
t^{(3)}_{0i}=(v^{(3)}_{\rm gr})_ik_0(1-\mathfrak{L}).\eend The
positive definiteness of the energy density (\ref{endensity})
results again in the requirement that the inequality (\ref{unicaus})
be satisfied. The causality in the form of the dominant energy
condition (\ref{poynting}) makes us expect that  vector
(\ref{components}) should be non-spacelike.
 Now, from (\ref{components}) with the use of the dispersion
law (\ref{linear3}) this condition becomes \bee\label{momsqr}
t_{0,\mu}^{(3)2}=k_0^2\{(k_3^2-k_0^2)(1-\mathfrak{L_F})^2
+k_\perp^2(1-\mathfrak{L_F}-2\mathfrak{FL_{FF}})^2\}=\nonumber\\= -
2\mathfrak{FL_{FF}}k_0^2k_\perp^2(1-\mathfrak{L_F}-2\mathfrak{FL_{FF}})
\leq 0.\eend Owing to relation (\ref{parallel}), this is exactly
equivalent to the requirement (\ref{group}) that the group velocity
of mode-3 photons should not exceed the speed of light in the
vacuum.

The same operations, performed over the energy-momentum tensor
(\ref{en-mom}) taken on mode 2, result (after omitting the positive
factor -$k\widetilde{F}^2k$) in an expression for the
energy-momentum tensor $t_{\mu\nu}^{(2)}$ that is obtained from
(\ref{T0mu}) by the duality transformation $F\to\widetilde{F}$,
$\mathfrak{L_{FF}}\to \mathfrak{L_{GG}}.$ When $\mathfrak{F}>0$, in
the special frame $(F^2k)_{1,2}=0,$
$(F^2k)_{0,3}=2\mathfrak{F}k_{0,3},$ so

\bee\label{components2}t_{0\nu}^{(2)}=k_0\left(\begin{tabular}{c}$k_0(1-\mathfrak{L_F}+2\mathfrak{FL_{GG}})
$\\$k_1(1-\mathfrak{L_F})$\\$k_2(1-\mathfrak{L_F}$
\\$k_3(1-\mathfrak{L_F}+2\mathfrak{FL_{GG}})$\end{tabular}\right)_\nu.\eend

The positivity of the energy density $t_{00}^{(2)}$ leads to the
inequality (\ref{unitarity2}). The group velocity of mode 2 is again
parallel to the momentum density 3-vector \bee\label{parallel2}
t^{(2)}_{0i}=(v^{(2)}_{\rm
gr})_ik_0(1-\mathfrak{L_F}+2\mathfrak{FL_{GG}}).\eend The causality
in the form of the dominant energy condition (\ref{poynting}) leads
from (\ref{components2}) with the use of the dispersion law
(\ref{linear2}) to \bee\label{momsqr2} t_{0\mu}^{(2)2} =
-2\mathfrak{FL_{GG}}k_0^2(k_0^2-k_3^2)(1-\mathfrak{L_F}+2\mathfrak{FL_{GG}})
\leq 0.\eend Owing to relation (\ref{parallel2}), this is exactly
equivalent to the requirement (\ref{group}) that the group velocity
of mode-2 photons should not exceed the speed of light in the
vacuum. Bearing in mind that eq. (\ref{unitarity2}) is already
established, eq. (\ref{min2}) follows from (\ref{momsqr2}).

To resume, we were able to reproduce in this Subsection the
requirements (\ref{unitarity2})--(\ref{min2}), but the remaining
requirements (\ref{unitarity3}) and (\ref{LGG}) do not follow from
(\ref{momsqr}), although the latter does not contradict them. Since,
as it was explained, the  form of the causality conditions
(\ref{poynting}) used in this Subsection is equivalent to the group
velocity restriction (\ref{group}), we think that our analysis has
indicated that the energy-density nonnegativity (\ref{endensity})
condition is somewhat weaker than the unitarity condition in the
form (\ref{unitarity}).

 The fulfillment of (\ref{momsqr}),  (\ref{momsqr2}) is
guaranteed by the inequalities (\ref{unitarity2}), (\ref{min2})--
(\ref{unitarity3}) established 
in Subsection B. However, the inverse statement would be wrong: the
inequalities (\ref{momsqr}), (\ref{momsqr2}), derived in the present
Subsection do not yet lead to (\ref{unitarity2}), (\ref{min2})--
(\ref{unitarity3}). This may indicate that pair of conditions
(\ref{unitarity}) (unitarity as the positivity of the residue) and
(\ref{group}) (causality as the boundedness of the group velocity),
used to derive the limitations (\ref{unitarity2}) --
(\ref{unitarity3}) of Subsection B, are together more restrictive
than the two principles (\ref{endensity}) (energy positiveness) and
(\ref{poynting}) (causality as non-spacelikeness of the
energy-momentum density), although the latter provide the fact that
when solving the Cauchy problem initial data have no influence on
what occurs outside their light cone. (This is proved in
\cite{hawking} within General Relativity context.)

\section{Testing certain Lagrangians}
\subsection{Euler-Heisenberg effective Lagrangian}
In the one-loop approximation of QED the quantities involved can be
calculated either using the Euler-Heisenberg effective Lagrangian
$\mathfrak{L}=\mathfrak{L^{(1)}}$ \cite{heisenberg}, when the
infrared limit is concerned,
 or, alternatively, the one-loop polarization operator calculated
in \cite{batalin} for off-shell photons -- within and beyond this
limit. In the infrared limit the photon-momentum-independent
coefficients in (\ref{kappa}) within one loop are the following
functions
 of the dimensionless magnetic field $b=eB/m^2$, where $e$ and $m$ are the electron
charge and mass:\bee\label{L_F}\mathfrak{L^{(1)}_F}
=\frac\alpha{2\pi}\int_0^\infty\frac{\rmd t}{t}\exp\left({-\frac
t{b}}\right)\left(\frac{-\coth\;t}{t}+\frac1{\sinh^2t}+\frac2{3}\right),\eend
\bee\label{L_GG}
2\mathfrak{F}\mathfrak{L^{(1)}_{GG}}=\frac\alpha{3\pi}\int_0^\infty\frac{\rmd
t}{t}\exp\left({-\frac
t{b}}\right)\left(\frac{-3\coth\;t}{2t}+\frac3{2\sinh^2t}+t\coth
t\right),\eend
\bee\label{L_FF}  2\mathfrak{F}\mathfrak{L^{(1)}_{FF}}=
\frac\alpha{2\pi}\int_0^\infty\frac{\rmd t}{t}\exp\left({-\frac
t{b}}\right) \left(\frac{\coth\;t}{t}- \frac{2t\coth t}{\sinh^2
t}+\frac1{\sinh^2t}\right).\eend Here $\alpha=e^2/4\pi=1/137$ is the
fine-structure constant. (We refer to the Heaviside-Lorentz system
of units with $c=\hbar=1$).    Eq. (\ref{L_F}) turns to zero as
$\mathfrak{F}\sim b^2$, since the divergent linear in $\mathfrak{F}$
part of the one-loop diagram was absorbed in the course of
renormalization into $\mathfrak{L}_{\rm cl}$. It can be verified
that the general relations (\ref{unitarity2})--(\ref{LGG}) ordained
by unitarity (\ref{unitarity}) and causality (\ref{causality}) to
the infrared limit  are obeyed by the one-loop approximation within
the vast range of the magnetic field values. (We are not considering
in the present context the electric-like case, since the (one-loop)
Heiseberg-Euler Lagrangian suffers the known instability under
spontaneous production of electron-positron pairs.)
 However, due to the
known lack of asymptotic freedom in QED \cite{ritus2}, some of the
general relations are violated for  exponentially strong fields of
Planck scale. One can establish the asymptotic behavior of
(\ref{L_F}) - (\ref{L_FF}) in the limit
$b=eB/m^2\rightarrow\infty$\bee\label{asymp}
\mathfrak{L_F^{(1)}}\simeq \frac\alpha{3\pi}(\ln b-1.79),\qquad
2\mathfrak{F}\mathfrak{L_{GG}^{(1)}}\simeq
\frac\alpha{3\pi}(b-1.90),\qquad
2\mathfrak{F}\mathfrak{L_{FF}^{(1)}}\simeq \frac\alpha{3\pi}.\eend
One can see then that the convexity properties
 (\ref{min2}), (\ref{LGG}) and hence the inequalities (\ref{epsil}), (\ref{mu})
 are left intact under arbitrarily strong magnetic field within one loop. So is the inequality (\ref{unitarity2}),
  thanks to the linearly growing \cite{skobelev} term in $\mathfrak{L_{GG}^{(1)}}$. On the contrary, eq.(\ref{unitarity3})
   is violated for $b>b^{\rm cr}_1=\exp\{0.79+3\pi/\alpha\}$, and eq.
   (\ref{unicaus}) for $b>b^{\rm cr}_2=\exp\{1.79+3\pi/\alpha\}>b_1^{\rm cr}.$

 Let us inspect  consequences of these violations.
 First note that the
inequality (\ref{causality}) requires  that $f_a(k_\perp^2)\geq 0$,
hence no branch of any dispersion curve may get into the region
$k_0^2-k_3^2<0$. If it might, the photon energy $k_0$ would have an
imaginary part within the momentum interval
$0<k_3^2<-f_a(k_\perp^2),$ corresponding to the vacuum excitation
exponentially growing in time. This sort of ghost would signal the
instability of the magnetized vacuum. Inequality (\ref{causality})
further requires that \bee\label{rest}\frac{\rmd
f_a^{\frac1{2}}(k_\perp^2)}{\rmd k_\perp}\leq 1,\qquad {\rm
or}\qquad f_a^{\frac1{2}}(k_\perp^2)\leq const+k_\perp.\eend  All
the dispersion curves (\ref{linear2}), (\ref{linear3}) in the
infrared approximation we are dealing with correspond to zero-mass
vacuum excitations $\left. k_0\right|_{k_3=k_\perp=0}=0$ -- photons,
since $f(0)=0$. Therefore \textit{const} = 0.

 Consider, first, mode 2. We mentioned that relation
(\ref{unitarity2}), which is
 the positive-norm condition for this mode,  is fulfilled for any large $b$. When
$b<b_2^{\rm cr},$  also the dispersion curve goes outside the light
cone, $\sqrt{k_0^2-k_3^3}\leq k_\perp$, as it is prescribed by eq.
(\ref{rest}) with $const =0$. However, the bracket in
(\ref{linear2}) becomes negative for $b>b_2^{\rm cr}$, and mode 2
becomes a complex energy ghost.

 Now comes  mode 3. The positive
norm condition for it, (relation  (\ref{unicaus})), is fulfilled,
when  $b<b_2^{\rm cr}$. However, within the range
 $b_1^{\rm cr}<b<b_2^{\rm cr}$ the bracket in (\ref{linear3}) is negative,
 and mode 3 is a complex energy ghost. For $b>b_2^{\rm cr}$ the dispersion curve (\ref{linear3})
 for mode-3 photon gets inside the light cone,$\sqrt{k_0^2-k_3^3}\geq k_\perp$, in contradiction with eq. (\ref{rest})
  and thus becomes a super-luminal excitation, tachyon, with real energy
 and negative norm. Note, that these superluminal excitations, peculiar to mode 3,  can hardly appear in
 reality, since the background field becomes unstable before
  it can reach, when growing, the necessary critical value $b=b_2^{\rm cr}$.
  An instability of the magnetized vacuum with respect to production of
 a constant field is associated with the imaginary energy at zero momentum. The elementary excitation
 with this property appears in mode 3 at a smaller threshold value, $b_3^{\rm cr}$,
 than in mode 2, $b_2^{\rm cr}$. The instability associated with mode-2 ghosts may
 lead to gaining the constant field with $\mathfrak{G}\neq 0$, since  the
 (pseudo)vector-potential $\flat^{(2)}_{\mu}$ (\ref{vectors}) carries
 an electric field component, parallel to the background magnetic
 field, whereas in $\flat^{(3)}_{\mu}$ this component is perpendicular to $\bf
 B$.

The borders of stability of the magnetic field found here by
 analyzing the one-loop approximation are characterized by the
 large exponential $\exp\{1/\alpha\}$. It is much larger than the
 border found earlier \cite{2006} as the value where the
 mass defect of the bound electron-positron pair completely
 compensates the 2$m$ energy gap between the electron and
 positron, which is of the order of $\exp\{1/\sqrt\alpha\}.$
\subsection{Born-Infeld Lagrangian}
 The situation is quite different for the Born-Infeld
 electrodynamics with its Lagrangian \bee\label{born} L_{\rm
 tot}=L^{\rm BI}=a^2\left(1-\sqrt{1+\frac{2\mathfrak{F}}{a^2}
 -\frac{\mathfrak{G}^2}{a^4}}\right)\eend
viewed upon as final, not
 subject to  further quantization. Here $a$ is an arbitrarily
 large  parameter with the dimensionality of mass squared. The
 correspondence principle (\ref{corresp}) is respected by eq. (\ref{born}).
 It does not contain field derivatives, hence all the infra-red limits encountered in this paper
 should be understood as exact values, for instance, going to the limit is unnecessary
 in (\ref{from17a}), (\ref{from17b}), (\ref{from17c}). The
 Lagrangian (\ref{born}) was derived long ago \cite{born} basing on very general
 geometrical principles of reparametrization-invariance, and
 besides it attracted much attention in recent decades thanks to the
 fact that it appears responsible for  the electromagnetic  sector of a
 string theory \cite{tseytlin} and thus is expected not to suffer from the lack of asymptotic freedom.
  For this reason our statement to follow that
 all the fundamental requirements established in Section 2 are
 obeyed in the Born-Infeld electrodynamics (\ref{born}) is
 instructive. We assume again that there is the constant and homogeneous
 magnetic-like external background and set
  $\mathfrak{G}=0$ after differentiation.
 Then, we get from
 (\ref{born})\bee\label{born2} 1-\mathfrak{L^{\rm
 BI}_F}=
 \left(1+\frac{2\mathfrak{F}}{a^2}\right)^{-\frac {1}{2}}\geq 0,\quad
 \mathfrak{L^{\rm BI}_{FF}}=a^{-2}
 \left(1+\frac{2\mathfrak{F}}{a^2}\right)^{-\frac3{2}}\geq 0,\quad \mathfrak{L^{\rm BI}_{GG}}=
 a^{-2}\left(1+\frac{2\mathfrak{F}}{a^2}\right)^{-\frac1{2}}\geq 0,\nonumber\\
 1-\mathfrak{L^{\rm BI}_F}+2\mathfrak{F}\mathfrak{L^{\rm BI}_{GG}}=
 \left(1+\frac{2\mathfrak{F}}{a^2}\right)^{\frac1{2}}\geq
0,\qquad 1-\mathfrak{L^{\rm BI}_F}-2\mathfrak{F}\mathfrak{L^{\rm
BI}_{FF}}=
 \left(1+\frac{2\mathfrak{F}}{a^2}\right)^{-\frac3{2}}\geq
0\qquad\qquad\eend where $\mathfrak{L^{\rm
 BI}}=L^{\rm
 BI}+2\mathfrak{F}.$ Thus, relations
 (\ref{unitarity2})--(\ref{LGG}) are all
 satisfied, hence there are neither ghosts, nor tachyons. The mode 1
 remains nonpropagating. As for modes 2 and 3, their dispersion
 curves coincide, since $f_2(k_\perp^2)=f_3(k_\perp^2)$ in (\ref{linear2}), (\ref{linear3}) due eqs.
 (\ref{born2}). This reflects the known absence of birefringence
 in the Born-Infeld electrodynamics \cite{plebanski}. Still,
 beyond the mass shell one has $\varkappa_2\neq\varkappa_3$, consequently the
 corresponding permeabilities  (\ref{from17a}), (\ref{from17b}), (\ref{from17c}) are different.
The same as in the one-loop QED, in the limit of large external
field there is a linearly growing contribution in $\varkappa_2$, so
mode 2 dominates, the dielectric permeability (\ref{from17b})
behaving like the middle equation in
(\ref{asymp})\bee\label{behaving}\varepsilon_{\rm long}^{\rm BI}(0)
\simeq 2\mathfrak{F}\mathfrak{L_{GG}^{\rm BI}}\simeq \frac B{a}\eend
with the identification $a=(3\pi/\alpha)B_0$, where $B_0=
m^2/e=4.4\times 10^{14}$ Gauss is the characteristic field strength
in QED. As a matter of fact, however, it is believed that $a$ should
be of the Planck scale $a\simeq m_{\rm Pl}^2/e=5.8\cdot 10^{44}B_0.$

If we include the electric-like case we shall see that eqs.
(\ref{born2}) are all fulfilled within the interval $-(a^2/2) <
\mathfrak{F} <\infty$, at the border of which the Lagrangian
(\ref{born}) becomes imaginary (recall that $\mathfrak{G}=0$.)
\subsection{Lagrangians giving rise to spontaneous magnetic field}
In this Subsection we consider, as counterexamples, two effective
Lagrangians that lead to nonzero magnetic field as the minimum
energy point and are thus conventionally interpreted as
spontaneously producing a constant homogeneous magnetic field
$B_{\rm sp}$.
 In  both of these cases below, one of which relating to a nonAbelian gauge theory, the fundamental properties of the Lagrangian
  established in Section IIB are violated in and around the point $B=B_{\rm sp}$.
\subsubsection{Batalin-Matinian-Savvidy Lagrangian}
These authors calculated \cite{bms} -- with the one-loop accuracy
and using Schwinger's proper-time method -- the effective Lagrangian
in the Yang-Mills theory as a function
 of two time- and space-independent field invariants.

 The intensity  tensor $G_{\mu\nu}^a=\partial A_\mu^a-\partial A_\nu^a-g\epsilon^{abc}A_\mu^bA_\nu^c$ is subject to the sourceless
 equation \bee
\label{sequation} \nabla^{ab}_\nu G^b_{\nu\mu}=0 \eend with the
standard covariant derivative
$\nabla^{ab}_\mu=\delta^{ab}\partial_\mu+gA_\mu^{ab},$
$A_{ab}=\epsilon^{acb}A_\mu^c$. Here the superscript $a$ is
responsible for the isotopic degree of freedom,  the subscript
$\mu=(i,0)$ runs the space-time  components,  $g$ is the coupling
constant,
 and $\epsilon^{abc}$ are the structural constants of SU(2). The simplest solution of the equation (\ref{sequation}) is the
  covariant constant field that satisfies the equation\bee
\label{ssequation} \nabla^{ab}_\rho G^b_{\nu\mu}=0. \eend It follows
from (\ref{ssequation}) that the intensity tensor factorizes as
$G_{\mu\nu}^a=F_{\mu\nu}n^a$, i.e. it is directed in the isotopic
space along a permanent direction of the constant (chosen as unit)
isotopic vector $n^a$, $F_{\mu\nu}$ being a constant tensor,
carrying the "chromomagnetic" and "chromoelectric"  background
fields. In a special gauge the vector potential may be chosen as
 $A_\mu^a=A_\mu n^a=-(1/2)F_{\mu\nu}x_\nu n^a$. It is seen that the present case is mostly close to quantum electrodynamics, the
 calculations can be made in a gauge-independent way and the result for the effective Lagrangian depends on the background Abelian
 field
  via the  field invariants $\mathfrak{F}$ and $\mathfrak{G}$ defined in terms of the tensor $F_{\mu\nu}$ in the same way  as in QED.

  The polarization operator responsible for propagation of small
  nonAbelian fields (gluons) against the background considered is, generally, defined by an equation
  similar to (\ref{Pi}) \bee\label{similar} \Pi_{\mu\tau}^{ab}(x,y)=
\left. \frac{\delta^2\Gamma}{\delta A_\mu^a(x)\delta A_\tau^b(y)}
\right |_{\mathfrak{G}=0,\; \mathfrak{F}= {\rm
const},\,A_\mu^a=A_\mu n^a }.\eend Then the polarization operator
(\ref{Pi}) is the projection of (\ref{similar}) to the only isotopic
direction
\bee\label{Pi2}\Pi_{\mu\tau}(x,y)=n^an^b\Pi_{\mu\tau}^{ab}(x,y)
.\eend This quantity governs the propagation of small perturbations
of the background field polarized  in the isotopic space parallel to
that field (call them chromophotons). The polarization operator
(\ref{Pi2}) possesses  all the properties exploited in Section II,
hence it makes sense  the inspect whether the
Batalin-Matinian-Savvidy Lagrangian obeys the properties
(\ref{unitarity2})--(\ref{LGG}) relating to propagation of long-wave
low frequency chromophotons.

  The total Lagrangian is again $L=-\mathfrak{F}+\mathfrak{L}$, where $-\mathfrak{F}$ is the tree Lagrangian on the covariantly
  constant fields under consideration. After renormalization the one-loop result of Ref. \cite{bms} for the real part of the
  effective Lagrangian
  $\mathfrak{L}$ can be represented as \bee\label{bms}
\mathfrak{L}(\mathfrak{F,G^2})=\widetilde{\mathfrak{L}}(\mathcal{B,E})=\frac1{8\pi^2}\int_0^\infty\frac{\rmd
s}{s}\left\{\frac{g^2\mathcal{B}\mathcal{E}}{\sinh(g\mathcal{B}s)
\sin(g\mathcal{E}s)}-\frac1{s^2}+
\frac{g^2(\mathcal{B}^2-\mathcal{E}^2)}{6} \right\}\rme^{-\mu^2s}+
\nonumber\\+\frac 1{4\pi^2}\int_0^\infty\frac{\rmd
s}{s}g^2\left\{\mathcal{E}\mathcal{B}\left[\frac{\sin(g\mathcal{B}s)}{\sinh(
g\mathcal{E}s)}-\frac{\sin(g\mathcal{E}s)}
{\sinh(g\mathcal{B}s)}\right]+\mathcal{E}^2-\mathcal{B}^2\right\}\rme^{-\mu^2s},\qquad\eend
where the invariant combinations $\mathcal{B}$ and $\mathcal{E}$ are
defined by (\ref{variables}) and coincide with the chromo-magnetic
and chromo-electric fields in a special Lorentz frame, respectively.
The normalization condition, obeyed by (\ref{bms}), contrary to
(\ref{corresp}), was imposed in a nonzero point
\bee\label{norm}\left.\frac{\rmd\mathfrak{L}(\mathfrak{F},0)}{\delta\mathfrak{F}}\right|_{\sqrt{2\mathfrak
{F}}=\mu^2}\equiv \left.\frac1{\mathcal{B}}\frac{\rmd
\mathfrak{\widetilde{L}}(\mathcal{B},0)}{\rmd
\mathcal{B}}\right|_{\mathcal{B}=\mu^2}=0. \eend The equality here
is the first line of (\ref{subst}). The integral in (\ref{bms}) is
convergent in the ultraviolet ($s\simeq0$) and the infrared
($s\simeq\infty)$ regions of the proper-time integration variable
$s$.

When $\mathfrak{G}=0$ and $\mathfrak{F}>0$, one has $\mathcal{E}=0$
and $\mathcal{B}=\sqrt{2\mathfrak{F}}$. \bee\label{bms0}
\mathfrak{L}(\mathfrak{F},0)=\widetilde{\mathfrak{L}}(\mathcal{B},0)=\frac1{8\pi^2}\int_0^\infty\frac{\rmd
s}{s}\left\{\frac{g\mathcal{B}}{s\sinh(g\mathcal{B}s)}-\frac1{s^2}+
\frac{g^2\mathcal{B}^2}{6} \right\}\rme^{-\mu^2s}+ \nonumber\\+\frac
1{4\pi^2}\int_0^\infty\frac{\rmd
s}{s}g^2\left\{\mathcal{B}\frac{\sin(g\mathcal{B}s)}{gs}
-\mathcal{B}^2\right\}\rme^{-\mu^2s},\qquad\eend

 The asymptotic behavior of (\ref{bms}) and of (\ref{bms0}) at $\mathfrak{F}\to\infty$ are the same as at $\mu^2\rightarrow0$,
 since (\ref{bms}) is a function of the ratio $\mu^2/\mathcal{B}$. Eq. (\ref{bms0}) behaves as
 \bee\label{asympt}\mathfrak{L}(\mathfrak{F},0)\asymp
 -\frac{11}{48\pi^2}g^2\mathfrak{F}\ln\left(\frac{2g^2\mathfrak{F}}{\mu^4}\right).\eend
Correspondingly, in the leading order \bee\label{L_f} \mathfrak{L_F}
=-\frac{11}{48\pi^2}g^2\ln\left(\frac{2g^2\mathfrak{F}}{\mu^4}\right),\qquad
2\mathfrak{FL_{FF}}=-\frac{11g^2}{24\pi^2}. \eend

It follows from (\ref{bms}) with the use of (\ref{subst})
that\bee\label{Lgg}
 2\mathfrak{F}\mathfrak{L_{GG}}
= \frac{1}{\mathcal{E}}\left. \left(\frac{\partial
\mathfrak{\widetilde{L}}(\mathcal{B},\mathcal{E})}{\partial
\mathcal{E}}\right)\right|_{\mathcal{E}=0}+
 \frac 1{\mathcal{B}}\frac{~\rmd \mathfrak{\widetilde{L}}(\mathcal{B},0)}{\rmd
 \mathcal{B}}=\nonumber\\
= \frac{g^2}{4\pi^2} \int_0^\infty\frac{\rmd t}{t}\left\{\frac
{-t\sin t}3 \right.+\left.\frac{\sinh t-t\cosh t} {2t\sinh^2 t}
 \right.+ \nonumber\\\\+\left.\frac{\sin t}{t}+\cos
t-\frac{11}{6}\frac{t}{\sinh
t}\right\}\exp\left(-\frac{\mu^2}{g\mathcal{B}}t\right),\eend where
$t=g\mathcal{B}s$. The integral of the first term in the bracket is
readily calculated to be equal to -1 in the limit
$(\mu^2/g\mathcal{B}) =0$, whereas the rest of it converges -- even
without the infrared regularization -- to a constant calculated
numerically. The convergence of (\ref{Lgg}) in the limit of infinite
magnetic field, unlike the QED expression (\ref{L_GG}), is the
formal reason why the linearly growing contribution to the
dielectric permeability of the magnetized vacuum, found responsible
for the formation of a string-like Coulomb potential in QED
\cite{2008}, is absent from chromomagnetized vacuum.  Finally, in
the above limit, we get\bee\label{lgg2}
2\mathfrak{F}\mathfrak{L_{GG}}=
-\frac{g^2}{4\pi^2}\left(\frac1{3}+1.5...\right)=-\frac{11g^2}{24\pi^2}.\eend
Contrary to (\ref{asymp}), the contribution, linear in the magnetic
field, is not present here.

We see from (\ref{L_f}), (\ref{lgg2}) that the general conditions
(\ref{unicaus}), (\ref{unitarity2}) and (\ref{unitarity3}), derived
in Sec. II for the dielectric and magnetic permeabilities, are
obeyed, while the convexity properties (\ref{min2}) and (\ref{LGG})
are not. So, the chromomagnetized vacuum is free, within the
one-loop approximation, of superluminal excitations and ghosts,
characteristic of the Euler-Hiesenberg approximation in QED, as
described in Subsection A above. On the contrary, the wrong
convexity $\mathfrak{L_{FF}}<0$ results in the fact that the
effective potential $V_{\rm eff}=-\mathfrak{L}$ has its minimum at a
nonvanishing value of the magnetic field \cite{savvidy}. Bearing in
mind that any constant magnetic field satisfies exact equation of
motion without sources due to gauge invariance, it is concluded that
the nonzero magnetic field is produced spontaneously. (As distinct
to the scalar Higgs case, the equation for potential minimum is not
an equation of motion for the gauge field.) However, the shift to
the minimum point does not result in improving the wrong convexity
sign. The matter is that there is an instability of the magnetic
field reflected in appearance of imaginary part of the effective
Lagrangian (already for magnetic-like case under consideration) due
to contribution of unstable gluon mode in a magnetic field
\cite{nielsen} into the spectral decomposition of the effective
action. (The presence of the imaginary part not seen in \cite{bms}
may be reproduced \cite{cabo2}, \cite{cabshab} also in calculations
following the Schwinger's proper time technique). This instability
is known to be resolved by going out of the sector of
covariantly-constant fields.

\subsubsection{Kawati-Kokado Lagrangian}
The Lagrangian of the named authors \cite{kawati} is remarkable in
that it proclaims  spontaneous production of the magnetic field as
large as 10$^{45} - 10^{47}$ G in the course of inflation. The model
includes  interaction between an electromagnetic and a complex
massless scalar fields considered in  de Sitter space-time. When
there is no direct coupling between the scalar field and the de
Sitter metric field, the  Lagrangian, calculated as a function of a
constant magnetic field, which  satisfies sourceless equations of
motion, is\bee\label{kawati} L=-\frac
1{2}B^2-\frac{e^2B^2}{192\pi^2}\left(\ln\frac{e^2B^2}{\kappa^2}+\alpha\right)+\frac{H^2\ln
2}{8\pi^2}|eB|,\eend where $H$ is the Hubble constant incorporated
in the de Sitter metric, $\kappa$ is a parameter taken to adjust the
dimension, and $\alpha$ is a certain numerical parameter. The
convexity of the Lagrangian (\ref{kawati}) with respect to the
variable $\mathfrak{F}=-B^2/2$ is upward in the region $\mathfrak
{F}>0$, in other words  condition (\ref{LGG}) is violated throughout
the magneticlike domain of $\mathfrak{F}$. As a consequence, the
effective potential, which is the Lagrangian taken with the opposite
sign, has a minimum at $B=B_{\rm sp}$ with \bee\label{minimum}
B_{\rm sp}=\frac{eH^2}{8\pi^2}.\eend (The small quantity $(e^2/192\;
\pi^2)$ was neglected.) The value of the spontaneous magnetic field
listed above is obtained in \cite{kawati} taking the typical values
for the Hubble constant, $H\sim 10^{15}-10^{17}$Gev, in
(\ref{minimum}). Its existence is completely due to the violation of
the general principles, reflected in eq. (\ref{LGG}). Note that, as
distinct from the Higgs mechanism, the wrong convexity of the
Lagrangian is not improved after the shift to the value
(\ref{minimum}). The other general requirement, eq. (\ref{unicaus}),
is violated for $B<B_{\rm sp}.$ Unlike the QED case of Subsection A,
this violation occurs at small values of the magnetic field. (We
cannot check conditions (\ref{unitarity2}) and (\ref{min2}), since
calculations with the second field-invariant $\mathfrak{G}$ kept
different from zero are not available.)

\subsection{Yang-mills field with external source}
The one-loop effective Lagrangian as a function of the background
Yang-Mills (gluon) field
 that has a nonvanishing classical source $J_\mu^a$ was calculated in \cite{cabshab}, \cite{cabo} within a special
 quantization procedure needed to substitute for the gauge invariance violated by that source. In this approach the vanishing of the covariant derivative
 $\nabla^{ac}_kJ^c_k(t)$, required by the gauge invariance, is achieved by treating this derivative as the secondary constraint. Correspondingly, under quantization, the functional delta-function $\delta(\nabla^{ac}_kJ^c_k(x))$ appears in the functional integral  over the gluon field to restrict, in the course of integration, their values involved in this covariant derivative.

  Let there be a constant background (classical) SU(2) Yang-Mills potential that in a special Lorentz frame and in a special gauge has the form\bee\label{y-mp}A_i^a=(A^2/3)^{1/2}\delta^a_i,\qquad A_0^a=0,\eend where $\delta^a_i$ is the Kronecker symbol and $A^2=A_\mu^aA_\mu^a$. Here the superscript $a$ is responsible for the isotopic degree of freedom, while the subscript $\mu=(i,0)$ marks  the space-time  components. The field intensity tensor of the constant potential (\ref{y-mp}) is $G_{\mu\nu}^a=g\epsilon^{abc}A_\mu^bA_\nu^c$, where $g$ is the selfcoupling constant, and $\epsilon^{abc}$ are the SU(2) fully antisymmetric unit tensor. The Yang-Mills equation is \bee\label{equation} \nabla^{ab}_\nu G^b_{\nu\mu}=-\frac 2{3}g^2A^2A_\mu^a, \eend with the standard covariant derivative  $\nabla^{ab}_\mu=\delta^{ab}\partial_\mu+gA_\mu^{ab},$ $A_{ab}=\epsilon^{acb}A_\mu^c$. We see that the constant field (\ref{y-mp}) requires the nonvanishing space-like current \bee\label{current} J_\mu^a=\frac 2{3}g^2A^2A_\mu^a\eend to be supported with. The classical field (\ref{y-mp}) obviously satisfies the current-conservation condition $\delta(\nabla^{ac}_kJ^c_k(x)=0.$ In what follows we use the notation for the field invariant $\mathfrak{F}=(1/4)G^a_{\mu\nu}G^a_{\mu\nu}.$ The normalization condition  $\left.{\rm d^4 Re}L/{\rm d}A^4\right|_{G_{(0)}}=-4g_r^2$ is imposed  in an arbitrary normalization point $G^a_{\mu\nu}=G^a_{(0)\mu\nu}$ to fix the renormalized coupling constant $g$. Here $L=-\mathfrak{F}+\mathfrak{L}$ is the full and $\mathfrak{L}$ the effective
Lagrangian, the tree Lagrangian being $-\mathfrak{F}$. According to
Ref.\cite{cabo}  the calculation within one-gluon-one-ghost loop
gives for the real part of the latter
$(\mathfrak{F}\gg\mathfrak{F}_0)$
 \bee\label{ymaction} {\rm Re}\mathfrak{L}=-\mathfrak{F}\frac{25g^2}{16\pi^2}+\frac{3g^2}{16\pi^2}
\mathfrak{F}\ln\frac{\mathfrak{F}}{\mathfrak{F}_0}.\eend (The
principle of correspondence realizes differently from QED: radiative
corrections contribute also into the part, linear in $\mathfrak{F}$,
since the normalization point $\mathfrak{F}_0$ is not zero.)

It is seen that the Lagrangian (\ref{ymaction}) is a convex function
of $\mathfrak{F},$ $\mathfrak{L_{FF}}=(3g^2/16\pi^2\mathfrak{F})>0$,
throughout the whole  magneticlike domain of validity
$\mathfrak{F}\gg\mathfrak{F}_0$, unlike the Matinian-Savvidy and
Kawati-Kokado Lagrangians considered in Subsections C, D.
Consequently, no constant magnetic field is spontaneously produced.
However, the presence of nonzero imaginary part of the Lagrangian of
Ref. \cite{cabo}, Im$\mathfrak{L}=-(12.15g^2/6\pi^2)\mathfrak{F}$,
makes the theory unstable under creation of gluonic tachyons. Unlike
the case of Subsection C, their spectra turn to zero in the
zero-momentum point (see \cite{cabo} for details), which explains,
why no constant field is gained in the present case. As for
condition (\ref{unicaus}), it is violated for
$\mathfrak{F}>\mathfrak{F}\exp (22+16\pi^2/3g^2)$. Therefore the
effective Lagrangian in the theory of Ref.\cite{cabo} is closer to
that of Euler-Heisenberg in what concerns its causal-unitarity
properties: condition (\ref{LGG}) is fulfilled for arbitrarily large
magnetic field, while condition (\ref{unicaus}) is violated in the
domain of exponentially large fields, which signifies the lack of
asymptotic freedom in the both theories. (We cannot check conditions
(\ref{unitarity2}) and (\ref{min2}), since calculations with the
second field-invariant kept different from zero are not available.)

\section{Discussion}
 In  the present
paper, for establishing obligatory properties of the effective
Lagrangian we exploited two general principles --  unitarity and
causality -- taken in the special form of the requirements of
nonnegativity of the residue (\ref{unitarity}) and of boundedness of
the group velocity (\ref{group}).  We feel it necessary to confront
this way of action  with other approaches.

Usually, consequences of causality and unitarity are discussed
referring to holomorphic properties of the polarization operator (or
of the  dielectric permittivity tensor) that follow from the
retardation of the linear response and are expressed -- after being
supplemented by certain postulates concerning the high-frequency
asymptotic conditions -- as the Kramers-Kronig (once-subtracted)
dispersion relations. Although the general proof of an analog of the
Kramers-Kronig relation in a background field is lacking from the
literature, for the magnetized vacuum the holomorphity of the
polarization operator eigenvalues $\varkappa_a$ in a cut complex
plane of the variable $(k_0^2-k_3^2)$ was established within the
one-loop approximation \cite{annphys}, \cite{shabtrudy}, the
probability of electron-positron pair creation by a photon making
the cut discontinuity. Nevertheless, as we could see in Section III
A, this approximation includes appearance of negative-norm ghosts
and tachyons in contradiction with causality and unitarity. Thus,
the knowledge of the holomorphic properties is not enough to be sure
that the causality and unitarity requirements have been exploited at
full.

More specifically the causality  is approached by referring to what
is called "causal propagation". Here the Hadamard's method
\cite{courant} of characteristic surface (the wave front), across
which the first derivative of the propagating solution may undergo a
discontinuity is used. The propagation is causal if the normal
vector to the
characteristic surface is time- or light-like. 
Once the coefficients in the differential equation responsible for
the wave propagation are restricted in such a way as to meet this
requirement, the wave front propagates exactly with the speed of
light $c$ = 1 \cite{brillouin} and should be equal to the phase
velocity taken at infinite value of the frequency according to the
Leontovich theorem \cite{mandelshtam}. (Note, however, that the
infinite-frequency limit cannot be covered by any finite-order
differential equation; on the contrary, when considering the general
case of non-polynomial dispersion the Schwinger-Dyson set of
equations should be taken seriously as integro-differential
equations). Certain conditions obtained in this way that should be
obeyed by the "structural function $H$", the knowing of which is
equivalent to the effective Lagrangian, may be found among numerous
relations in a scrupulous study of Jerzy Pleba\'{n}ski. It seems,
however, that inequalities (9.176) derived in his Lectures
\cite{plebanski}, relating to the general case $\mathfrak{F}\neq0$,
$\mathfrak{G}\neq0,$ and the subsequent formulae, relating to the
 null-field subcase,
$\mathfrak{F}=\mathfrak{G}=0,$ need to be supplemented by
consequences of some requirements intended to substitute for
unitarity or positiveness of the energy, not exploited in
\cite{plebanski}, before/in-order-that a comparison with our
conclusions might become possible. In the case of nontrivial
dispersion, however, a coincidence is not even to be expected. The
point is that the requirement that the wave front should not
propagate faster than light is only a necessary, but not yet
sufficient condition of the causal propagation: other signals should
not be faster than light, either. It is widely recognized
\cite{bornwolf} that the group velocity is the speed of the wave
packet at least where no anomalous dispersion is present, in which
case the group velocity loses its interpretation as the wave packet
speed and may exceed unity. An extension of the group velocity into
the domain of anomalous dispersion that keeps it below the speed of
light is also possible \cite{newphysrev}. In  Ref. \cite{newphysrev}
we also argued why the excess of the group velocity over the speed
of light encountered in some problems with a violation of the
Lorentz invariance should be viewed upon as a serious discrepancy
with the relativity principle, understood in this case as
equivalence of a given reference frame, in which an external agent
like a background field is also present, with another inertial
frame, in which there is the same external agent, but
Lorentz-boosted from the initial frame.

This is why we treat the group velocity criterion as the causality
criterion in the present paper as well as in  \cite{newphysrev}.
Previously the appeal to the group
 velocity has shown its fruitfulness in establishing the phenomenon
 of canalization of the photon energy along the external magnetic
 field \cite{nuovocim}, \cite{annphys} and the capture of
 gamma-quanta by a strong nonhomogeneous magnetic field of a pulsar
 \cite{nature}, \cite{ass}. As for the violation of the group
 velocity criterion for exponentially strong magnetic field discovered for
 the one-loop approximation in Section
 III, we admitted that the necessary value of the  magnetic field cannot be achieved, because
 the magnetic field becomes unstable already at  smaller values.
 Therefore, a magnetic field higher than that, for which the photon
 may become superluminal, is to be ruled out like people use to  rule out perfectly
 elastic body, although in the latter case no mechanism that would ban
 its formation is  considered.

 On the other hand, the fulfillment  of the
 Dominant Energy Condition (DEC) (\ref{poynting})  implies that
 the causality is reassured, because
 when solving the Cauchy problem initial data have no influence on
what occurs outside their light cone. (This is proved in
\cite{hawking} within General Relativity context.) We saw in
Subection II D that the group-velocity criterion is
 equivalent to DEC in what concerns the consequences for the effective Lagrangian as a function
 of constant magnetic-like  background field, although
 the implementation of DEC and WEC to the problem of
 elementary excitations over the magnetized vacuum undertaken in Subsection C of Section
 II has indicated, however, as we already discussed it in that subsection, that these
two conditions together lead to somewhat weaker conclusions than the
ones that followed in Subsection B from
 imposing the conditions of unitarity in the form (\ref{unitarity}) and causality in the form(\ref{group}).

This work was supported by the Russian Foundation for Basic Research
(Project No. 05-02-17217), as well as by the Israel Science
Foundation of the Israel Academy of Sciences and Humanities.

\end{document}